\documentstyle[11pt]{article}
\begin{document}

\title{\textbf{QCD modified ghost scalar field dark energy models}}


\author{K. Karami$^{1,2}$\thanks{E-mail: KKarami@uok.ac.ir} ,
S. Asadzadeh$^{1}$, M. Mousivand$^{1}$, Z. Safari$^{1}$\\
$^{1}$\small{Department of Physics, University of Kurdistan,
Pasdaran St., Sanandaj, Iran}\\$^{2}$\small{Center for Excellence in
Astronomy \& Astrophysics of Iran (CEAAI-RIAAM), Maragha, Iran}}

\maketitle

\begin{abstract}
Within the framework of FRW cosmology, we study the QCD modified
ghost scalar field models of dark energy in the presence of both
interaction and viscosity. For a spatially non-flat FRW universe
containing modified ghost dark energy (MGDE) and dark matter, we
obtain the equation of state of MGDE, the deceleration parameter as
well as a differential equation governing the MGDE density
parameter. We also investigate the growth of structure formation for
our model in a linear perturbation regime. Furthermore, we
reconstruct both the dynamics and potentials of the quintessence,
tachyon, K-essence and dilaton scalar field DE models according to
the evolution of the MGDE density.
\end{abstract}

\clearpage
\section{Introduction}
Astronomical data from the supernova type Ia (SNeIa)
\cite{1,1b,1c,1d,1e}, cosmic microwave background (CMB)
\cite{2,2b,2c}, large scale structure (LSS) \cite{3,3b,3c}, baryon
acoustic oscillations (BAO) \cite{4} and weak lensing \cite{5}
indicate that expansion of the universe is speeding up rather than
decelerating. It is believed that the present accelerated expansion
is driven by gravitationally repulsive dominant energy component
known as ``dark energy'' (DE). Although the nature of DE remains a
mystery, various models of DE have been proposed in the literature
(for review see \cite{Padmanabhan,Padmanabhanb,Padmanabhanc}).
Theoretically, the simplest candidate for such a component is a
small positive cosmological constant, but it suffers the
difficulties associated with the fine tuning and the cosmic
coincidence problems \cite{Weinberg}.

More recently, a new DE model called ghost DE (GDE) has been
motivated from the Veneziano ghost of choromodynamics (QCD)
\cite{Urban,Urbanb,Urbanc}. The Veneziano ghost is required to exist
for the resolution of the $U(1)_A$ problem in low energy effective
theory of QCD. The ghosts make no contribution in the flat Minkowski
space, but once they are in the curved space or time-dependent
background such as our Friedmann-Robertson-Walker (FRW) universe,
the cancelation of their contribution to the vacuum energy is
off-set, leaving a small energy density $\sim\Lambda^3_{\rm QCD}H$,
where $H$ is the Hubble parameter and $\Lambda_{\rm QCD}\sim100$ MeV
is the QCD mass scale \cite{Witten,Wittenb,Wittenc,Wittend,Wittene}.
This small contribution can play an important role in the
evolutionary behavior of the universe. For instance, taking
$H\sim10^{-33}$eV at the present, $\Lambda^3_{\rm QCD}H$ gives the
right order of observed magnitude of the DE density. This
coincidence is remarkable and implies that the GDE model gets rid of
fine tuning problem \cite{Urban,Urbanb,Urbanc}. In addition, the
appearance of the QCD scale could be relevant for a solution to the
cosmic coincidence problem, as it may be the scale at which dark
matter (DM) forms \cite{Forbes}. The other advantage of the GDE with
respect to other DE models include the fact that it can be
completely explained within the standard model and general
relativity (GR), without recourse to any new field, new degree(s) of
freedom, new symmetries or modifications of GR. It is worth to note
that the GDE model does not violate unitarity, causality, gauge
invariance and other important features of renormalizable quantum
field theory, as advocated in
\cite{Zhitnitsky1,Zhitnitsky1b,Zhitnitsky1c}.

The ordinary GDE density is given by \cite{Urban,Urbanb,Urbanc}
\begin{equation}
\rho_{D}=\alpha H,\label{GDE}
\end{equation}
where $\alpha$ is a constant with dimension $[{\rm energy}]^3$, and
roughly of order of $\Lambda_{\rm QCD}^3$. This new kind of DE model
has got a lot of enthusiasm recently in the literature
\cite{Cai,sheikh,sheikhb,sheikhc,sheikhd,sheikhe,sheikhf,Sheykhi1,Sheykhi2,Sheykhi3,Rozas-Fernandez}.
Although the ordinary GDE model is consistent with the observational
data, it suffers from the difficulty to describe the early evolution
of the universe. This motivated Cai et al. \cite{Cai2} to introduce
the modified GDE (MGDE) density as
\begin{equation}\label{rho}
    \rho_{D}=\alpha H+\beta H^{2},
\end{equation}
where the constant $\alpha$ is same as that defined in the ordinary
GDE and $\beta$ is another constant with dimension $[{\rm
energy}]^2$. Note that the Veneziano ghost field in QCD is
originated from the vacuum energy of quantum fields which is of the
form $H+{\mathcal O}(H^2)$ \cite{Zhitnitsky2}. However, in the
ordinary GDE model, only the leading term $H$ has been considered.
Although the subleading term $H^2$ is too small and cannot drive the
universe to accelerating expansion, Cai et al. \cite{Cai2} showed
that this term can play an important role in the early evolution of
the universe, acting as an early DE. Using the joint analysis of the
astronomical data, Cai et al. \cite{Cai2} found that the subleading
term $H^2$ of the DE density--the early DE--could have a fraction
energy density around $10\%$. Indeed, the term proportional to $H^2$
is related to the difference between the vacuum energies in
Minkowski space and in a FRW universe \cite{Maggiore,Maggioreb}. On
the other hand, the vacuum energy difference from the Veneziano
ghost field introduced in order to solve the so-called $U(1)_A$
problem in QCD has the exact form $\alpha H+\beta H^{2}$, where
$\alpha\sim\Lambda_{\rm QCD}^3\sim (100~{\rm MeV})^3$. For other
motivations to consider this form of DE, see again \cite{Cai2},
where the MGDE model has been fitted with current observational data
including SNeIa, BAO, CMB, big bang nucleosynthesis (BBN), Hubble
parameter and growth rate of matter perturbation in order to get
some constraints on the model parameters. It was found that the MGDE
model, without having the two fundamental cosmological puzzles, like
the $\Lambda$CDM fit the astronomical data very well.

On the other hand, as is well known, an alternative proposal for DE
is scalar field scenarios such as quintessence, tachyon, K-essence
and dilaton (for review see \cite{Copeland} and references therein).
The scalar field models are an effective description of an
underlying theory of DE and can alleviate the fine tuning and cosmic
coincidence problems \cite{Ali}. Scalar fields naturally arise in
particle physics including supersymmetric field theories and
string/M theory. Therefore, scalar field is expected to reveal the
dynamical mechanism and the nature of DE. Although fundamental
theories such as string/M theory do provide a number of possible
candidates for scalar fields, they do not uniquely predict their
field $\phi$ and potential $V(\phi)$ \cite{JPWu}. Therefore it
becomes meaningful to reconstruct both the dynamics and potentials
of the scalar fields from some DE models such as holographic
\cite{Zhang11,Zhang11b,Zhang11c,Zhang11d,Zhang11e,Zhang11f,Zhang11g},
agegraphic \cite{Zhang15,Zhang15b,Zhang15c,Zhang15d} and ordinary
ghost \cite{Sheykhi2,Sheykhi3,Rozas-Fernandez}.

Describing the DE model in a scalar field framework provides a more
fundamental representation of the dark component. This motivates us
to establish different scalar field models according to evolutionary
behavior of the MGDE scenario. To do so, in section 2 we investigate
the MGDE in a spatially non-flat FRW universe. In section 3 we study
the growth of structure formation in our model. In section 4 we
reconstruct both the dynamics $\phi$ and potentials $V(\phi)$ of the
quintessence, tachyon, K-essence and dilaton scalar field models of
DE according to the evolution of MGDE density. Section 5 is devoted
to conclusions.
\section{The Veneziano MGDE and DM}

Within the framework of Einstein gravity, we consider a spatially
non-flat FRW universe filled with MGDE density $\rho_D$ and DM
energy density $\rho_{\rm m}$. Therefore, the first Friedmann
equation reads
\begin{equation}
H^2+\frac{k}{a^2}=\frac{1}{3M_P^2}~ (\rho_{D}+\rho_{\rm
m}),\label{eqfr}
\end{equation}
where the scalar curvature $k=0,1,-1$ denote a flat, closed and open
FRW universe, respectively. Also $M_{P}=(8\pi G)^{-1/2}$ is the
reduced Planck mass.

Using the fractional energy densities
\begin{equation}
\Omega_{\rm m}=\frac{\rho_{\rm m}}{\rho_{\rm cr}}=\frac{\rho_{\rm
m}}{3M_P^2H^2},~~~\Omega_{D}=\frac{\rho_{D}}{\rho_{\rm
cr}}=\frac{\rho_{D}}{3M_P^2H^2},~~~\Omega_{k}=\frac{k}{a^2H^2},
\label{eqomega}
\end{equation}
the Friedmann equation (\ref{eqfr}) can be rewritten as
\begin{equation}
1+\Omega_{k}=\Omega_{D}+\Omega_{\rm m}.\label{eq10}
\end{equation}
Substituting Eq. (\ref{rho}) into the middle relation of Eq.
(\ref{eqomega}) gives
\begin{equation}\label{OmegaDMGDE}
 \Omega_{D}=\frac{\alpha}{3M^{2}_{P}H}+1-\gamma,
\end{equation}
where
\begin{equation}
\gamma=1-\frac{\beta}{3M_{P}^{2}}.
\end{equation}
Using Eq. (\ref{OmegaDMGDE}), the curvature energy density parameter
takes the form
\begin{equation}\label{omegak}
    \Omega_{k}=\frac{\Omega_{k_0}}{(\Omega_{D_0}+\gamma-1)^{2}}\left(\frac{\Omega_{D}+\gamma-1}{a}\right)^2,
\end{equation}
where we take $a_0=1$ for the present time and the subscript ``0''
denotes the present value of a quantity.

Following the observational evidences we extend our study to the
case in which the MGDE has a bulk viscosity property
\cite{Zimdahl,Zimdahlb,Zimdahlc} and also interact with DM
\cite{Bertolami,Bertolamib}. It is well known that in the framework
of FRW metric, the shear viscosity has no contribution in the energy
momentum tensor, and the bulk viscosity behaves like an effective
pressure. Because, the CMB does not indicate significant anisotropy
due to shear viscosity and only bulk viscosity is taken into account
for the fluids in the cosmological context \cite{Jaffe}. In viscous
cosmology, shear and bulk viscosities arise in relation to space
anisotropy and isotropy, respectively. It was also pointed out that
the bulk viscosity can play a significant role in the formation of
the LSS of the universe \cite{Folomeev}. It can also alleviate
several cosmological puzzles like age problem and cosmic coincidence
problem \cite{Brevik,Brevikb,Brevikc,Brevikd}.

The interaction between DE and DM can be detected in the formation
of LSS. It was suggested that the dynamical equilibrium of collapsed
structures such as galaxy clusters would be modified due to the
coupling between DE and DM. The recent observational evidence
provided by the galaxy cluster Abell A586 supports the interaction
between DE and DM \cite{Bertolami8}. The other observational
signatures on the dark sectors' mutual interaction can be found in
the probes of the cosmic expansion history by using the SNeIa, BAO
and CMB shift data \cite{Honorez}.

In the presence of bulk viscosity and interaction, the energy
densities of MGDE and DM do not conserve separately and continuity
equations take the forms
\begin{equation}
\dot{\rho}_{D}+3H(1+\omega_{D})\rho_{D}=9H^{2}\xi-Q,\label{eqpol}
\end{equation}
\begin{equation}
\dot{\rho}_{\rm m}+3H\rho_{\rm m}=Q,\label{eqCDM}
\end{equation}
where $\omega_{D}=p_{D}/\rho_{D}$ is the equation of state (EoS)
parameter of MGDE. Also $\xi=\epsilon H^{-1}\rho_{D}$ is the bulk
viscosity coefficient with the viscosity constant $\epsilon$
\cite{Brevik,Brevikb,Brevikc,Brevikd} and
$Q=3b^{2}H(\rho_{D}+\rho_{\rm m})$ is the interaction term with the
coupling constant $b^2$ \cite{Kim06}. This expression for the
interaction term $Q$ was first introduced in the study of the
suitable coupling between a quintessence scalar field and a
pressureless cold DM field \cite{Amendola1}.

Taking time derivative of Eq. (\ref{eqfr}) and using Eqs.
(\ref{eqomega}), (\ref{eq10}), (\ref{eqpol}) and (\ref{eqCDM}) one
can get
\begin{equation}
\frac{\dot{H}}{H^2}=-\frac{3}{2}\left[1+\frac{\Omega_k}{3}+(\omega_{D}-3\epsilon)\Omega_{D}\right].\label{HdotH2}
\end{equation}
Taking time derivative of Eq. (\ref{rho}), using (\ref{HdotH2}), and
substituting the obtained result into Eq. (\ref{eqpol}) gives the
EoS parameter of the interacting viscous MGDE as
\begin{equation}\label{omega}
    \omega_{D}=\frac{1}{\Omega_{D}-\gamma-1}
    \left[1-\frac{\Omega_{k}}{3}+\frac{\gamma-1}{\Omega_{D}}\left(1+\frac{\Omega_{k}}{3}\right)+2b^{2}
    \left(\frac{1+\Omega_{k}}{\Omega_{D}}\right)\right]+3\epsilon.
\end{equation}
For completeness, we give the deceleration parameter
\begin{equation}
q=-1-\frac{\dot{H}}{H^2},
\end{equation}
which combined with the dimensionless density parameter $\Omega_D$
and the EoS parameter $\omega_D$ form a set of useful parameters for
the description of the astrophysical observations. Replacing Eq.
(\ref{omega}) into (\ref{HdotH2}) gives
\begin{equation}
q=\frac{\gamma+2(\Omega_{D}-1)-\Omega_k+3b^2(1+\Omega_k)}{\Omega_{D}-\gamma-1}
  .\label{q}
\end{equation}
Taking the derivative of Eq. (\ref{OmegaDMGDE}) with respect to
redshift $z=\frac{1}{a}-1$ and using Eqs. (\ref{HdotH2}) and
(\ref{omega}) gives
\begin{equation}\label{omegaprim}
\frac{{\rm d}\Omega_{D}}{{\rm
d}z}=-\left(\frac{3}{1+z}\right)\left(\frac{\Omega_{D}+\gamma-1}{\Omega_{D}-\gamma-1}\right)
\left[\Omega_{D}-1-\frac{\Omega_{k}}{3}+b^2(1+\Omega_{k})\right].
\end{equation}
Note that Eqs. (\ref{q}) and (\ref{omegaprim}) show that both the
deceleration and MGDE density parameters are independent of
viscosity constant. The viscosity appeared only in the dynamical EoS
parameter of MGDE, Eq. (\ref{omega}).

\begin{figure}
\includegraphics{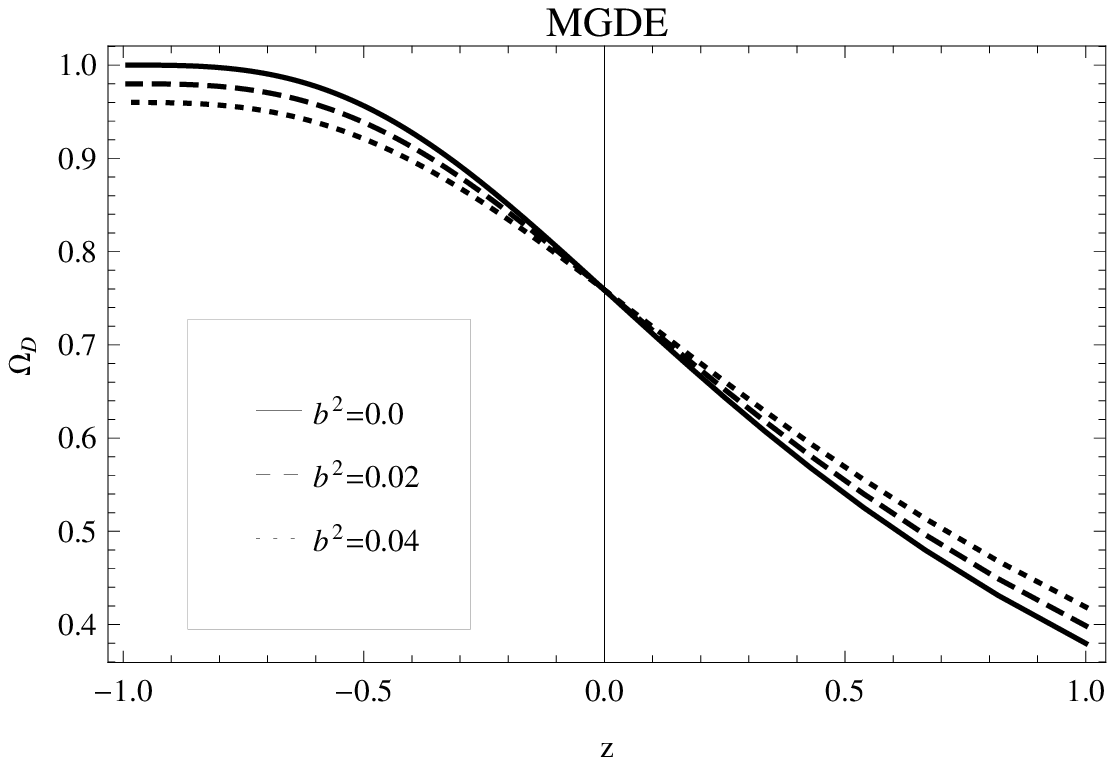}
      \vspace{4.7cm}
      \caption[]{MGDE density parameter, Eq. (\ref{omegaprim}), versus redshift for
      different coupling constants $b^2$.
      Auxiliary parameters are $\Omega_{k_0}=0.01$, $\Omega_{D_0}=0.76$ and $\gamma=1.105$ \cite{Cai2}.}
         \label{OmegaD-z-b2}
   \end{figure}

We solve the differential equation governing the MGDE density
parameter, Eq. (\ref{omegaprim}), numerically. In Fig.
\ref{OmegaD-z-b2}, variation of the MGDE density parameter
$\Omega_{D}$ versus redshift $z=\frac{1}{a}-1$ for different
coupling constants $b^2$ is plotted. Figure \ref{OmegaD-z-b2} shows
that: i) for a given $b^2$, $\Omega_D$ increases during history of
the universe. ii) At early and late times, $\Omega_D$ increases and
decreases, respectively, with increasing $b^2$.

\begin{figure}
\includegraphics{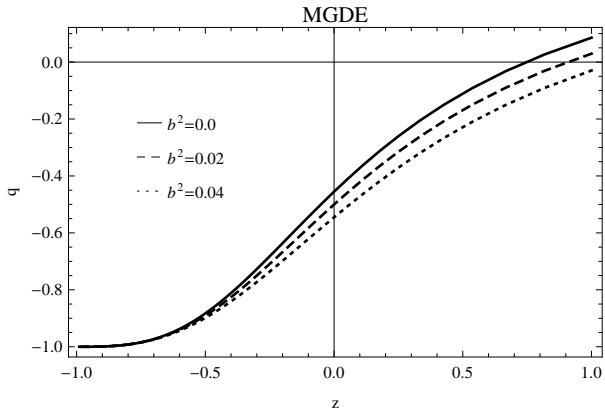}
      \vspace{4.7cm}
      \caption[]{Deceleration parameter, Eq. (\ref{q}), versus redshift for
      different coupling constants $b^2$. Auxiliary parameters as in Fig. \ref{OmegaD-z-b2}.}
         \label{q-z-b2e}
   \end{figure}

In Fig. \ref{q-z-b2e} we plot the evolutionary behavior of the
deceleration parameter, Eq. (\ref{q}), for different $b^2$. Figure
\ref{q-z-b2e} presents that: i) the universe transitions from a
matter dominated epoch at early times to the de Sitter phase, i.e.
$q=-1$, in the future, as expected. The result for $b^2=0$ is in
agreement with that obtained by \cite{Cai2}. ii) For $b^2=0.0$,
$0.02$ and $0.04$ at $z=0.75$, $0.91$ and $1.10$, respectively, we
have a cosmic deceleration $q>0$ to acceleration $q<0$ transition
which is compatible with the observations \cite{Ishida}. iii) For a
given $z$, although $q$ decreases with increasing $b^2$, in the far
future $q$ becomes independent of $b^2$.

\begin{figure}
\includegraphics{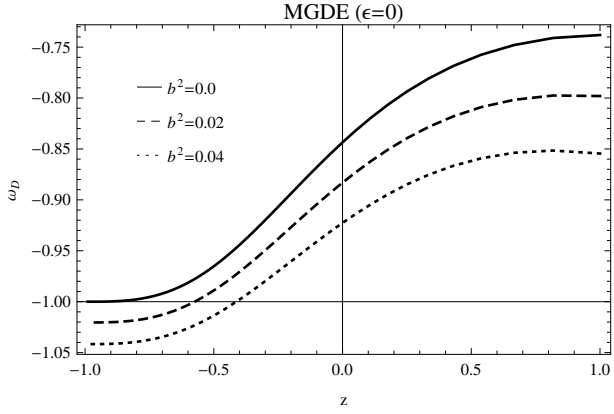}
      \vspace{4.7cm}
      \caption[]{EoS parameter of MGDE, Eq. (\ref{omega}), versus redshift for different
      coupling constants $b^2$ with $\epsilon=0$. Auxiliary parameters as in Fig. \ref{OmegaD-z-b2}.}
         \label{wD-z-b2}
   \end{figure}

\begin{figure}
\includegraphics{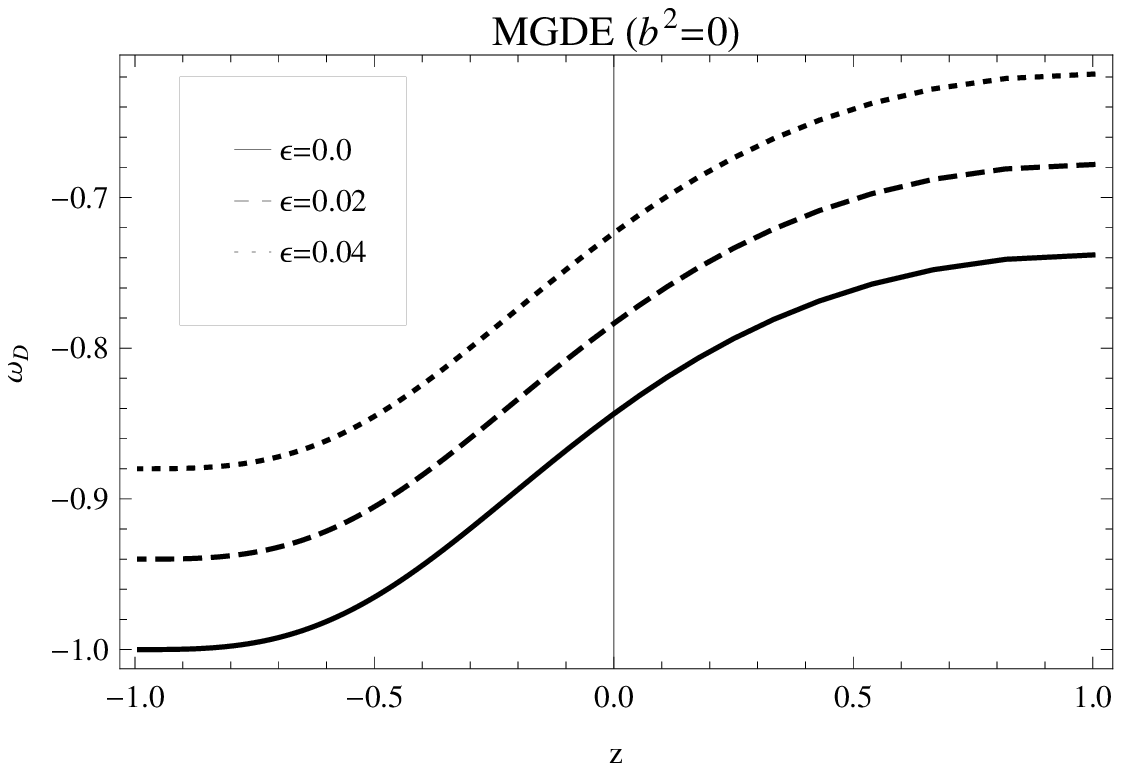}
      \vspace{4.7cm}
      \caption[]{Same as Fig. \ref{wD-z-b2} for different viscosity constants $\epsilon$ with $b^2=0$.}
         \label{wD-z-e}
   \end{figure}

\begin{figure}
\includegraphics{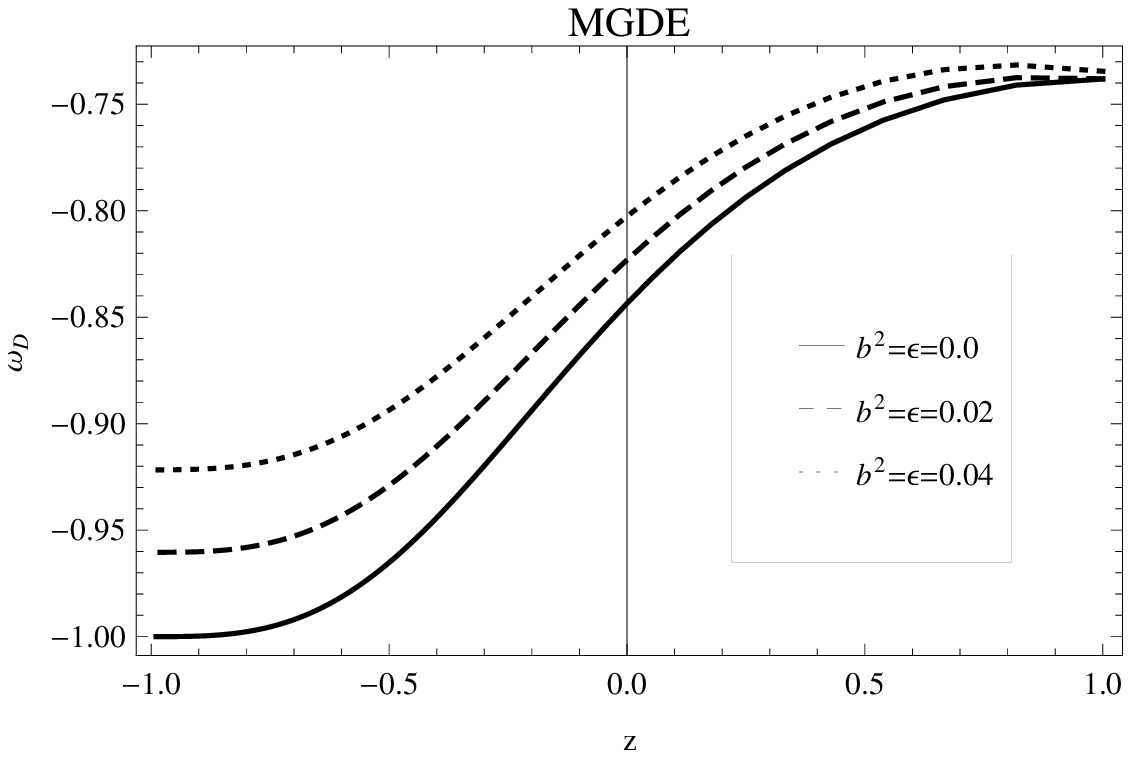}
      \vspace{4.7cm}
      \caption[]{Same as Fig. \ref{wD-z-b2} for different coupling $b^2$ and viscosity $\epsilon$ constants.}
         \label{wD-z-b2e}
   \end{figure}

The variation of the EoS parameter of interacting viscous MGDE, Eq.
(\ref{omega}), for different coupling $b^2$ and viscosity $\epsilon$
constants is plotted in Figs. \ref{wD-z-b2}, \ref{wD-z-e} and
\ref{wD-z-b2e}. Figures \ref{wD-z-b2} and \ref{wD-z-e} illustrate
that: i) in the absence of viscosity ($\epsilon=0$), $\omega_D(z)$
varies from the quintessence phases ($\omega_D>-1$) to the phantom
regime ($\omega_D<-1$). ii) In the absence of interaction ($b^2=0$),
$\omega_D(z)$ varies from $\omega_D>-1$ to $\omega_D=-1$, which is
similar to the freezing quintessence model \cite{Caldwell2}. iii)
For a given $z$, $\omega_D$ decreases and increases with increasing
$b^2$ and $\epsilon$, respectively. Note that the result for
$b^2=\epsilon=0$ is in accordance with that obtained by \cite{Cai2}.
Figures \ref{wD-z-b2e} shows that the interaction and viscosity have
opposite effects on $\omega_D$.
\section{The growth of structure formation}
Here we investigate the growth rate of matter in the presence of
interaction between viscous MGDE and DM. Following \cite{book}, the
structure formation takes place in the Newtonian regime. Hence, the
equations of motion containing the Euler and Poisson equations for
DM in the non-relativistic approximation reduce to
\begin{equation}
\frac{\partial {\mathbf v}}{\partial t}+({\mathbf v}.\nabla){
\mathbf v}=-\frac{\nabla p_{\rm m}}{\rho_{\rm
m}}-\nabla\phi,\label{eqm1}
\end{equation}
\begin{equation}
\nabla^2\phi=4\pi G\rho_{\rm m},\label{eqm2}
\end{equation}
where ${\mathbf v}$ and $\phi$ are the velocity of DM and
gravitational potential, respectively. According to the linear
perturbation theory \cite{book}, substituting the perturbations
\begin{equation}
{\mathbf v}\rightarrow{\mathbf v}+\delta{\mathbf v},~~~p_{\rm
m}\rightarrow p_{\rm m}+\delta p_{\rm m},~~~\rho_{\rm
m}\rightarrow\rho_{\rm m}+\delta\rho_{\rm
m},~~~\phi\rightarrow\phi+\delta\phi,
\end{equation}
into Eqs. (\ref{eqm1}) and (\ref{eqm2}) and retaining only first
order corrections to the background variable, one can obtain
\begin{equation}
\delta\dot{H}=-2H\delta H-\frac{4\pi G}{3}\delta\rho_{\rm
m},\label{pert1}
\end{equation}
where we have used $\nabla.{\mathbf v}=3H$ and set $v_s^2=0$ for the
pressureless DM ($p_{\rm m}=0$).

Now we generalize the linear perturbation formalism introduced in
\cite{book} to the case of interaction between viscous MGDE and DM.
Note that Eq. (\ref{pert1}) still holds in the presence of
interaction and viscosity. Because the bulk viscous coefficient
$\epsilon$ and the coupling constant of interaction term $b^2$ only
appear in the continuity Eqs. (\ref{eqpol}) and (\ref{eqCDM}).

From Eqs. (\ref{rho}) and (\ref{eqCDM}), we have
\begin{equation}
\dot{\rho}_{\rm m}+3H\rho_{\rm m}=3b^2H(\alpha H+\beta H^2+\rho_{\rm
m}).\label{rhok1}
\end{equation}
Substituting the perturbations $\rho_{\rm m}\rightarrow\rho_{\rm
m}+\delta\rho_{\rm m}$ and $H\rightarrow H+\delta H$ into the above
relation yields the density perturbation equation for DM as
\begin{equation}
\delta\dot{\rho}_{\rm m}+3H(1-b^2)\delta\rho_{\rm
m}=\left[\frac{\rho_{\rm m}}{H}+3b^2(\alpha H+2\beta
H^2)\right]\delta H.\label{deltarho}
\end{equation}
In terms of the DM density contrast $\delta_{\rm m}=\delta\rho_{\rm
m}/\rho_{\rm m}$, Eq. (\ref{deltarho}) can be rewritten for $\delta
H$ as
\begin{equation}
\delta H=\frac{\dot{\delta}_{\rm m}+\frac{3b^2}{\rho_{\rm m}}(\alpha
H^2+\beta H^3)\delta_{\rm m}}{-3(1-b^2)+\frac{3b^2}{\rho_{\rm
m}}(2\alpha H+3\beta H^2)}.\label{deltaH}
\end{equation}
Taking the time derivative of Eq. (\ref{deltaH}) gives
\begin{equation}
\delta\dot{H}=\frac{{\rm I}}{3\left[1-2b^2\Big(1+\frac{2\alpha
H+3\beta H^2}{\rho_{\rm m}}\Big)\right]},\label{deltaHdot}
\end{equation}
where
\begin{eqnarray}
{\rm I}&=&\left[b^2\left(1+\frac{2\alpha H+3\beta H^2}{\rho_{\rm
m}}\right)-1\right]\ddot{\delta}_{\rm m} \nonumber\\&
+&\frac{b^2}{\rho_{\rm m}}\left[-7\alpha H^2-6\beta H^3+\frac{8\pi
G}{3}\rho_{\rm m}(\alpha+3\beta H)\right]\dot{\delta}_{\rm m}
\nonumber\\& +&\frac{b^2}{\rho_{\rm m}}\left[-3\alpha H^3+4\pi
G\rho_{\rm m}\Big(2\alpha H+3\beta H^2\Big)\right]\delta_{\rm m},
\end{eqnarray}
and from Eq. (\ref{pert1}) we have used $\dot{H}=-H^2-4\pi
G\rho_{\rm m}/3$. We also have neglected the terms higher than
${\mathcal O}(b^2)$. Using the latest observations (golden SNeIa,
the shift parameter of CMB and the BAO) and combining them with the
lookback time data we have that $b^2$ could be as large as 0.2 (see
\cite{Feng}) although a value of $b^2<0.04$ is favored.

Inserting Eqs. (\ref{deltaH}) and (\ref{deltaHdot}) into
(\ref{pert1}) and using $\delta\rho_{\rm m}=\rho_{\rm m}\delta_{\rm
m}$, one can get the evolution equation for the dimensionless DM
density perturbation $\delta_{\rm m}$ as
\begin{eqnarray}
&&~~\left\{1-b^2\left(1+\frac{2\alpha H+3\beta H^2}{\rho_{\rm
m}}\right)\right\}\ddot{\delta}_{\rm m} \nonumber\\&&
+\left\{2H+b^2\left(\frac{3\alpha H^2}{\rho_{\rm m}}-\frac{8\pi
G}{3}(\alpha+3\beta H)-2H\right)\right\}\dot{\delta}_{\rm m}
\nonumber\\&& -\left\{4\pi G\rho_{\rm
m}-b^2\left[3H^3\left(\frac{3\alpha+2\beta H}{\rho_{\rm
m}}\right)+4\pi G\Big(2\alpha H+3\beta H^2+2\rho_{\rm
m}\Big)\right]\right\}\delta_{\rm m}=0.\label{deltarhom}
\end{eqnarray}
Note that the viscosity constant $\epsilon$ does not appear in Eq.
(\ref{deltarhom}). This comes back to the fact that in our model,
the DM has not the viscosity properties (see Eq. \ref{eqCDM}) and
due to choosing a specific form for the bulk viscosity coefficient
$\xi=\epsilon H^{-1}\rho_{D}$ in Eq. (\ref{eqpol}), the viscosity
constant $\epsilon$ does not affect $\delta_{\rm m}$ in our model.

In the absence of interaction, i.e. $b^2=0$, Eq. (\ref{deltarhom})
recovers the well known relation given by \cite{book}
\begin{equation}
\ddot{\delta}_{\rm m}+2H\dot{\delta}_{\rm m}-4\pi G\rho_{\rm
m}\delta_{\rm m}=0.\label{deltamb20}
\end{equation}
Using the following definitions
\begin{equation}
\bar{\ddot{\delta}}_{\rm m}=\ddot{\delta}_{\rm
m}/H_0^2,~~~\bar{\dot{\delta}}_{\rm m}=\dot{\delta}_{\rm
m}/H_0,~~~\bar{H}=H/H_0,~~~\bar{\alpha}=\frac{\alpha}{3M_P^2H_0},\label{dimenlesspara}
\end{equation}
one can rewrite Eq. (\ref{deltarhom}) in dimensionless form as
\begin{equation}
{\mathcal{J}}_1\bar{\ddot{\delta}}_{\rm
m}+{\mathcal{J}}_2\bar{\dot{\delta}}_{\rm
m}+{\mathcal{J}}_3\delta_{\rm m}=0,\label{delta}
\end{equation}
where
\begin{equation}
{\mathcal{J}}_1=1-b^2\left[1+\frac{2\bar{\alpha}/\bar{H}+3(1-\gamma)}{\Omega_{\rm
m}}\right],
\end{equation}
\begin{equation}
{\mathcal{J}}_2=2\bar{H}+b^2\left[\bar{\alpha}\left(\frac{3}{\Omega_{\rm
m}}-1\right)+(3\gamma-5)\bar{H}\right],
\end{equation}
\begin{equation}
{\mathcal{J}}_3=-\frac{3}{2}\Omega_{\rm
m}\bar{H}^2+3b^2\bar{H}\left[\left(\frac{3\bar{\alpha}+2(1-\gamma)\bar{H}}{\Omega_{\rm
m}}\right)+\bar{\alpha}+\frac{3}{2}(1-\gamma)\bar{H}+\Omega_{\rm
m}\bar{H}\right].
\end{equation}
In terms of the growth factor $f$ defined as \cite{book}
\begin{equation}
f(z)=\frac{{\rm d}\ln \delta_{\rm m}}{{\rm d}\ln a}=-(1+z)\frac{{\rm
d}\ln \delta_{\rm m}}{{\rm d}z},\label{f}
\end{equation}
equation (\ref{delta}) can be rewritten as
\begin{equation}
-(1+z)\bar{H}^2\frac{{\rm d}f}{{\rm
d}z}+\bar{H}\left[-(1+z)\frac{{\rm d}\bar{H}}{{\rm
d}z}+\bar{H}f+\left(\frac{{\mathcal{J}}_2}{{\mathcal{J}}_1}\right)\right]f+\left(\frac{{\mathcal{J}}_3}{{\mathcal{J}}_1}\right)=0.\label{fdiff}
\end{equation}
Note that for the matter dominated universe, i.e. $H^2=\rho_{\rm
m}/3M_P^2$, solution of Eq. (\ref{deltamb20}) yields $\delta_{\rm
m}=a$. In this case, Eq. (\ref{f}) gives the growth factor $f=1$.

\begin{figure}
\includegraphics{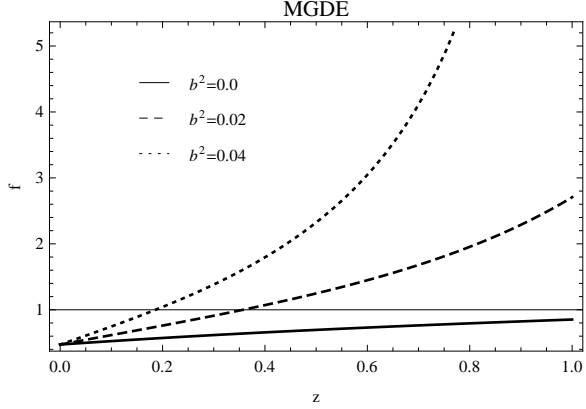}
      \vspace{4.7cm}
      \caption[]{Growth factor of DM, Eq. (\ref{fdiff}), versus redshift for
      different coupling constants $b^2$. Auxiliary parameters are
      $\Omega_{k_0}=0.01$, $\Omega_{D_0}=0.76$, $\gamma=1.105$ \cite{Cai2} and $f_0=0.473$ \cite{Cai2}. Legend as in Fig. \ref{OmegaD-z-b2}.}
         \label{f-z-b2}
   \end{figure}

In general, the differential equation (\ref{fdiff}) has no
analytical solution. Hence, we need to solve it numerically. To do
so, we first need to know the dimensionless Hubble parameter
$\bar{H}(z)$. From Eq. (\ref{OmegaDMGDE}) and last relation in Eq.
(\ref{dimenlesspara}) we obtain
\begin{equation}
\bar{H}(z)=\frac{\bar{\alpha}}{\Omega_{D}(z)+\gamma-1},
\end{equation}
and
\begin{equation}
\bar{\alpha}=\Omega_{D_0}+\gamma-1.
\end{equation}
Now with the help of $\Omega_D(z)$ plotted in Fig. \ref{OmegaD-z-b2}
which has been already obtained by numerically solving Eq.
(\ref{omegaprim}), one can obtain the evolutionary behavior of the
growth factor $f(z)$ of DM. In Fig. \ref{f-z-b2}, variation of the
growth factor $f(z)$ versus redshift for different coupling
constants $b^2$ is plotted. Figure \ref{f-z-b2} shows that: i) for a
given $b^2$, $f$ decreases during history of the universe. ii) For a
given $z$, $f$ increases with increasing $b^2$. The result for
$b^2=0$ is same as that obtained by \cite{Cai2}.

\begin{figure}
\includegraphics{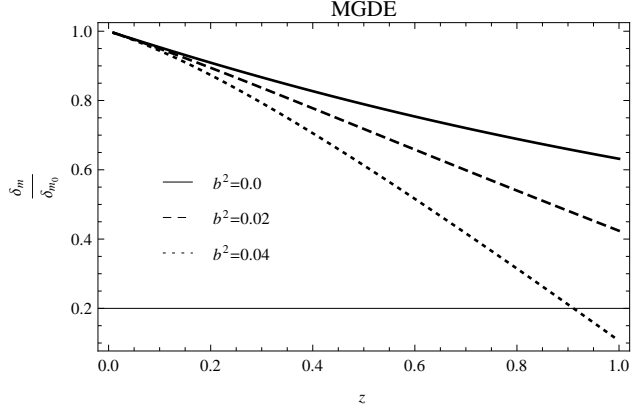}
      \vspace{4.7cm}
      \caption[]{Dimensionless DM
density perturbation, Eq. (\ref{deltamz}), versus redshift for
      different coupling constants $b^2$. Legend and auxiliary parameters as in Fig. \ref{f-z-b2}.}
         \label{delta-z-b2}
   \end{figure}

\begin{figure}
\includegraphics{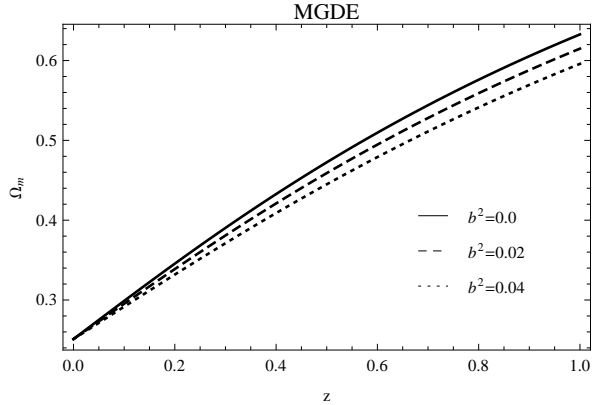}
      \vspace{4.7cm}
      \caption[]{DM density parameter, $\Omega_{\rm m}=1+\Omega_k-\Omega_D$, versus redshift for
      different coupling constants $b^2$. Legend and auxiliary parameters as in Fig. \ref{OmegaD-z-b2}.}
         \label{OmegaM-z-b2}
   \end{figure}

Using Eq. (\ref{f}), the dimensionless DM density perturbation
$\delta_{\rm m}$ takes the form
\begin{equation}
\delta_{\rm m}(z)=\delta_{\rm
m_0}\exp{\left[-\int_0^z\frac{f(z)}{1+z}~{\rm d
}z\right]}.\label{deltamz}
\end{equation}
In Fig. \ref{delta-z-b2}, we plot the evolutionary behavior of
$\delta_{\rm m}(z)$ for different $b^2$. Figure \ref{delta-z-b2}
presents that: i) for a given $b^2$, $\delta_{\rm m}$ increases
during history of the universe. ii) For a given $z$, $\delta_{\rm
m}$ decreases with increasing $b^2$. This shows a suppression of
structure growth relative to the non-interacting case. In the
presence of interaction, there is less DM in the past (see Fig. \ref
{OmegaM-z-b2}), and this leads to a suppression in the growth of
structure. This is in agreement with the result obtained by
\cite{Caldera}. Note that the suppression is specific to the
circumstance that the interacting and non-interacting cases are
normalized to have the same parameters today, i.e. $\Omega_{\rm
m_0}$ and $\delta_{\rm m_0}$.

\section{Correspondence with scalar field models}

Here, our aim is to investigate whether a minimally coupled scalar
field with a specific action/Lagrangian can mimic the dynamics of
the Veneziano MGDE model so that this model can be related to some
fundamental theory (such as string/M theory), as it is for a scalar
field. For this task, it is then meaningful to reconstruct the
dynamics $\phi$ and potential $V(\phi)$ of a scalar field model
possessing some significant features of the underlying theory of DE,
such as the MGDE model. To do so and following the method proposed
by \cite{Starobinsky}, we establish a correspondence between the
MGDE and various scalar field models by identifying their respective
energy densities and equations of state and then reconstruct both
the dynamics and potential of the field.

\subsection{Quintessence MGDE}

\begin{figure}
\includegraphics{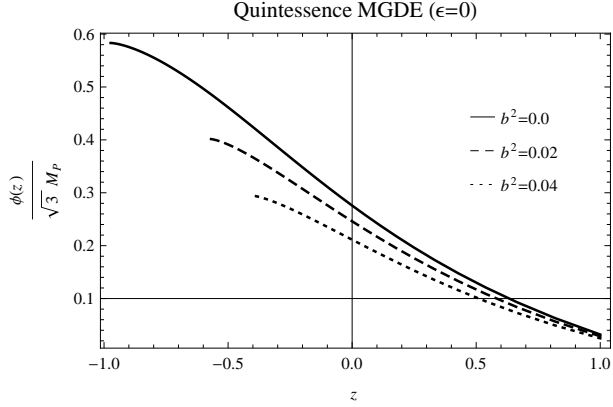}
      \vspace{4.7cm}
      \caption[]{Quintessence MGDE scalar field, Eq. (\ref{phiQ}), versus redshift for different coupling constants $b^2$ with $\epsilon=0$.
      Auxiliary parameters are $\Omega_{k_0}=0.01$, $\Omega_{D_0}=0.76$, $\gamma=1.105$ \cite{Cai2} and $\phi(1)=0$.}
         \label{PhiQ-z-b2}
   \end{figure}

\begin{figure}
\includegraphics{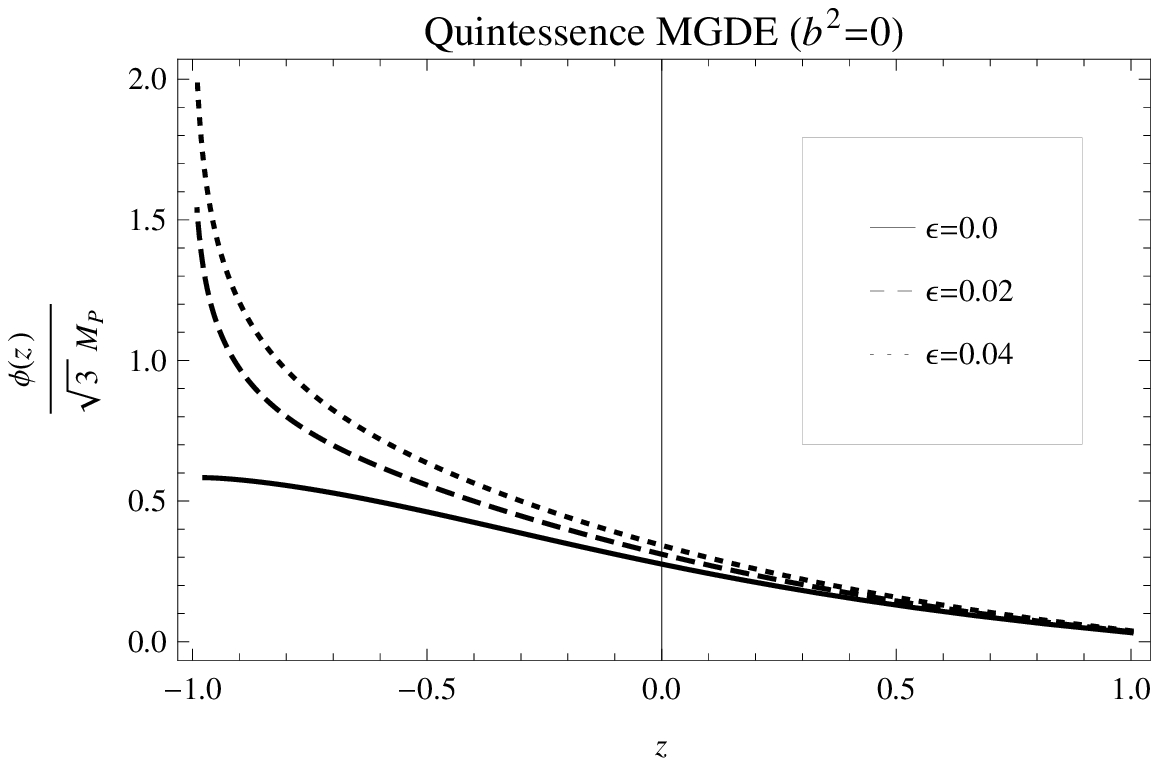}
      \vspace{4.7cm}
      \caption[]{Same as Fig. \ref{PhiQ-z-b2} for different viscosity constants $\epsilon$ with $b^2=0$.}
         \label{PhiQ-z-e}
   \end{figure}

\begin{figure}
\includegraphics{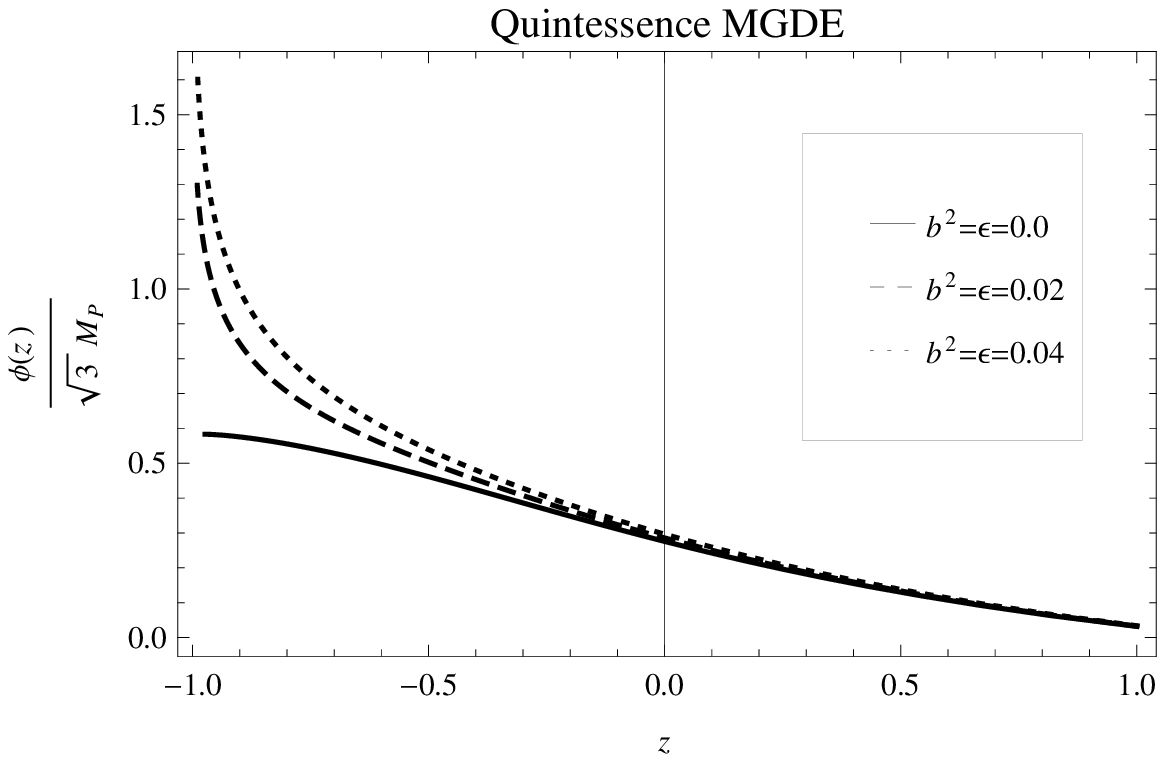}
      \vspace{4.7cm}
      \caption[]{Same as Fig. \ref{PhiQ-z-b2} for different coupling $b^2$ and viscosity $\epsilon$ constants.}
         \label{PhiQ-z-b2e}
   \end{figure}

 Quintessence is described by an
ordinary time dependent and homogeneous scalar field $\phi$ which is
minimally coupled to gravity, but with a particular potential
$V(\phi)$ that leads to the accelerating universe. The action for
quintessence is given by \cite{Copeland,Ratra,Ratrab}
\begin{equation}\label{1}
S=\int {\rm d}^4x\sqrt{-g}\left[
-\frac{1}{2}g^{\mu\nu}\partial_\mu\phi\partial_\nu\phi-V(\phi)
\right].
\end{equation}
The energy density and pressure of the quintessence scalar field
$\phi$ are as follows
\begin{equation}
\rho_Q=\frac{1}{2}\dot \phi^2+V(\phi),\label{rhoQ}
\end{equation}
\begin{equation}
p_Q=\frac{1}{2}\dot \phi^2-V(\phi).\label{p q}
\end{equation}
The quintessence EoS parameter takes the form
\begin{equation}
\omega_Q=\frac{p_Q}{\rho_Q}=\frac{\dot \phi^2-2V(\phi)}{\dot
\phi^2+2V(\phi)}.\label{omegaQ}
\end{equation}
Identifying Eq. (\ref{omegaQ}) with the EoS parameter of interacting
viscous MGDE (\ref{omega}), $\omega_Q=\omega_D$, and also equating
Eq. (\ref{rhoQ}) with (\ref{rho}), $\rho_Q=\rho_D$, one can get
\begin{equation}
V(\phi)=\frac{1}{2}(1-\omega_D)\rho_D,\label{Vphi-2}
\end{equation}
\begin{equation} \dot
\phi^2=(1+\omega_D)\rho_D.\label{phidot2-2}
\end{equation}
Inserting Eqs. (\ref{rho}) and (\ref{omega}) into the above
equations, one can get the quintessence potential and kinetic energy
as
\begin{eqnarray}
V(\phi)=\frac{\alpha^{2}\Omega_{D}}{6M^{2}_{P}(\Omega_{D}+\gamma-1)^{2}}
\left[\frac{\Omega_{D}-\gamma-2+\frac{\Omega_{k}}{3}-\left(\frac{\gamma-1}{\Omega_{D}}\right)\left(1+\frac{\Omega_{k}}{3}\right)-2b^{2}
    \left(\frac{1+\Omega_{k}}{\Omega_{D}}\right)}{\Omega_{D}-\gamma-1}
    -3\epsilon\right],\label{VphiQ}
\end{eqnarray}
\begin{eqnarray}
\dot\phi^2=\frac{\alpha^{2}\Omega_{D}}{3M^{2}_{P}(\Omega_{D}+\gamma-1)^{2}}
\left[\frac{\Omega_{D}-\gamma-\frac{\Omega_{k}}{3}+\left(\frac{\gamma-1}{\Omega_{D}}\right)\left(1+\frac{\Omega_{k}}{3}\right)+2b^{2}
    \left(\frac{1+\Omega_{k}}{\Omega_{D}}\right)}{\Omega_{D}-\gamma-1}
    +3\epsilon\right].\label{phidot2Q}
\end{eqnarray}
Integrating Eq. (\ref{phidot2Q}) with respect to $a$ and using
(\ref{OmegaDMGDE}) yields the modified ghost quintessence scalar
field as
\begin{eqnarray}
\phi(a)-\phi(a_i)=\sqrt{3}~M_{P}\int^{a}_{a_i}\sqrt{\Omega_{D}}
\left[\frac{\Omega_{D}-\gamma-\frac{\Omega_{k}}{3}+\left(\frac{\gamma-1}{\Omega_{D}}\right)\left(1+\frac{\Omega_{k}}{3}\right)+2b^{2}
    \left(\frac{1+\Omega_{k}}{\Omega_{D}}\right)}{\Omega_{D}-\gamma-1}
    +3\epsilon\right]^{\frac{1}{2}}
~\frac{{\rm d}a}{a}.\label{phiQ}
\end{eqnarray}
The above integral cannot be taken analytically. But with the help
of Eq. (\ref{omegak}) and numerical solution of the differential
equation (\ref{omegaprim}) one can obtain the evolutionary behavior
of the quintessence MGDE scalar field. In Figs. \ref{PhiQ-z-b2},
\ref{PhiQ-z-e} and \ref{PhiQ-z-b2e} we plot the variation of scalar
field, Eq. (\ref{phiQ}), versus redshift. Figures \ref{PhiQ-z-b2}
and \ref{PhiQ-z-e} present that: i) for a given $b^2$ or $\epsilon$,
$\phi$ increases during history of the universe. ii) For a given
$z$, $\phi$ decreases and increases with increasing $b^2$ and
$\epsilon$, respectively. Figure \ref{PhiQ-z-b2} shows that for
$b^2=0.0$, $0.02$ and $0.04$ at $z<-0.97$, $-0.57$ and $-0.41$,
respectively, $\phi$ becomes pure imaginary, i.e. $\dot{\phi}^2<0$,
and does not show itself in Fig. \ref{PhiQ-z-b2}. For
$\dot{\phi}^2<0$ the MGDE scalar field behaves as a phantom-type
scalar field \cite{Caldwell}. Figure \ref{PhiQ-z-b2e} clarifies that
interaction and viscosity have opposite effects on $\phi$. For
$b^2=\epsilon$, the effect of viscosity is more than the
interaction.

\begin{figure}
\includegraphics{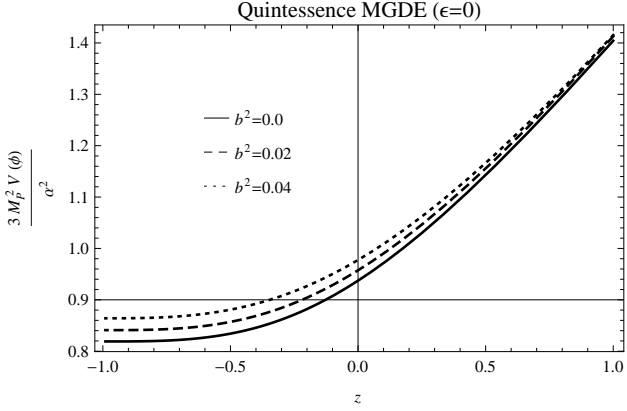}
      \vspace{4.7cm}
      \caption[]{Quintessence MGDE potential, Eq. (\ref{VphiQ}), versus redshift for different coupling constants $b^2$ with $\epsilon=0$.
      Auxiliary parameter as in Fig. \ref{PhiQ-z-b2}.}
         \label{VQ-z-b2}
   \end{figure}

\begin{figure}
\includegraphics{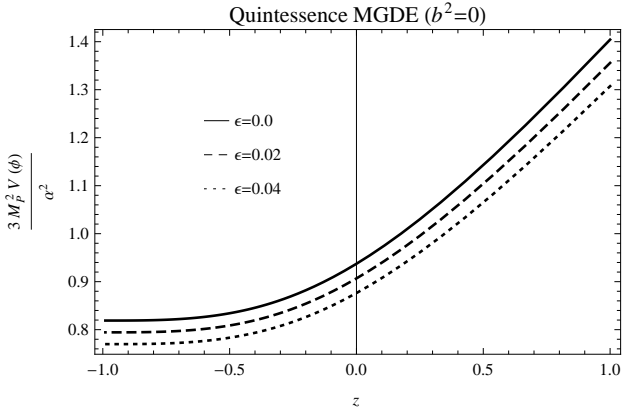}
      \vspace{4.7cm}
      \caption[]{Same as Fig. \ref{VQ-z-b2} for different viscosity constants $\epsilon$ with $b^2=0$.
      Auxiliary parameters as in Fig. \ref{PhiQ-z-b2}.}
         \label{VQ-z-e}
   \end{figure}

\begin{figure}
\includegraphics{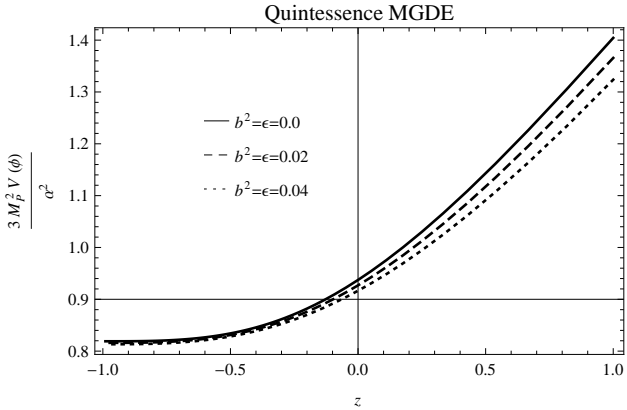}
      \vspace{4.7cm}
      \caption[]{Same as Fig. \ref{VQ-z-b2} for different coupling $b^2$ and viscosity $\epsilon$ constants.
      Auxiliary parameters as in Fig. \ref{PhiQ-z-b2}.}
         \label{VQ-z-b2e}
   \end{figure}

In Figs. \ref{VQ-z-b2}, \ref{VQ-z-e} and \ref{VQ-z-b2e}, the
variation of quintessence MGDE potential, Eq. (\ref{VphiQ}), versus
redshift is presented for different $b^2$ and $\epsilon$. Figures
\ref{VQ-z-b2} and \ref{VQ-z-e} illustrate that: i) for a given $b^2$
or $\epsilon$, the quintessence MGDE potential $V(\phi)$ decreases
during history of the universe. ii) For a given $z$, $V(\phi)$
increases and decreases with increasing $b^2$ and $\epsilon$,
respectively. Figure \ref{VQ-z-b2e} shows that in the presence of
both interaction and viscosity, although for $b^2=\epsilon$ at early
times the effect of viscosity is more than the interaction, at late
times they neutralize the effect of each other.

\subsection{Tachyon MGDE}

\begin{figure}
\includegraphics{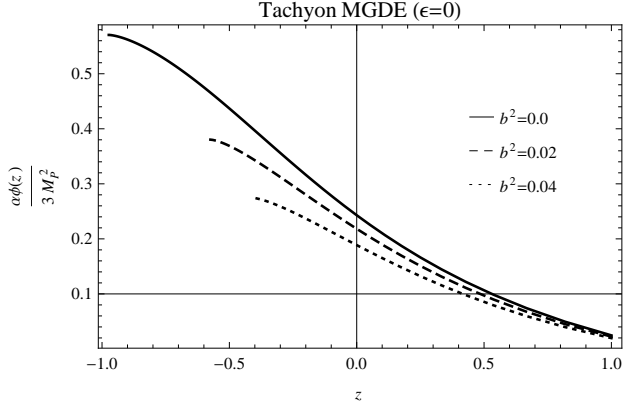}
      \vspace{4.7cm}
      \caption[]{Tachyon MGDE scalar field, Eq. (\ref{phiT}), versus redshift for different coupling constants $b^2$ with $\epsilon=0$.
      Auxiliary parameters as in Fig. \ref{PhiQ-z-b2}.}
         \label{PhiT-z-b2}
   \end{figure}

\begin{figure}
\includegraphics{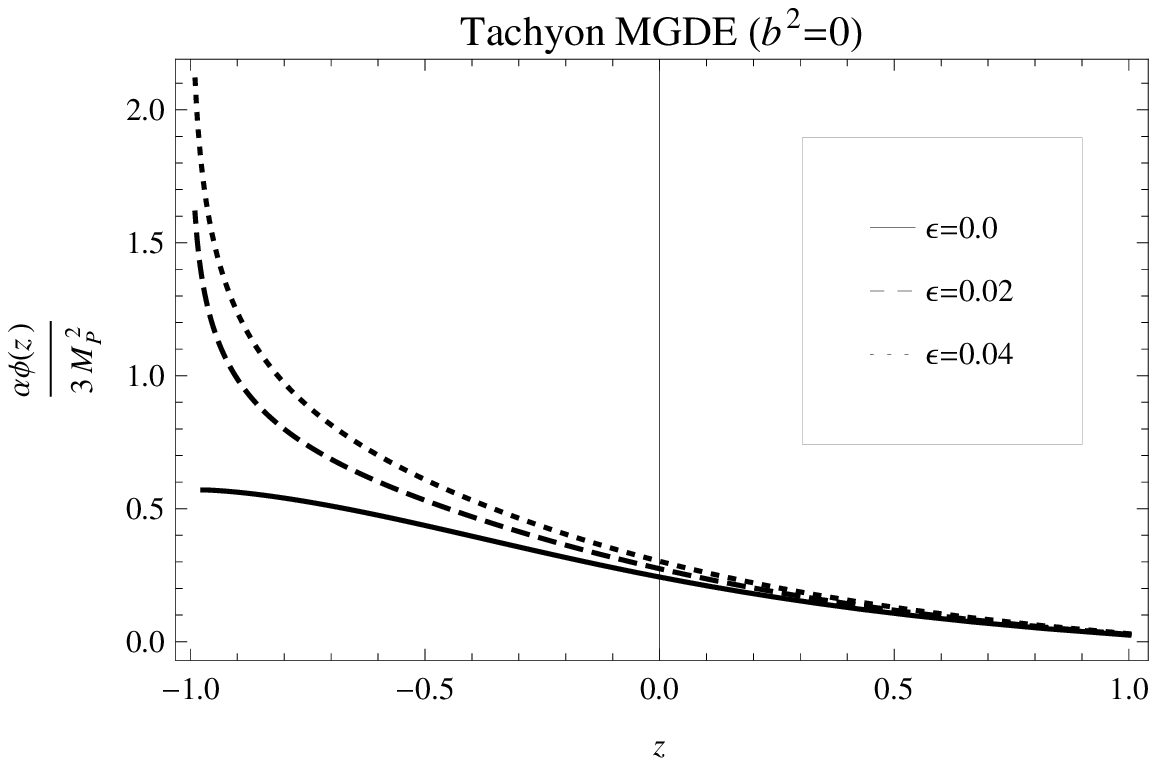}
      \vspace{4.7cm}
      \caption[]{Same as Fig. \ref{PhiT-z-b2} for different viscosity constants $\epsilon$ with $b^2=0$.}
         \label{PhiT-z-e}
   \end{figure}

\begin{figure}
\includegraphics{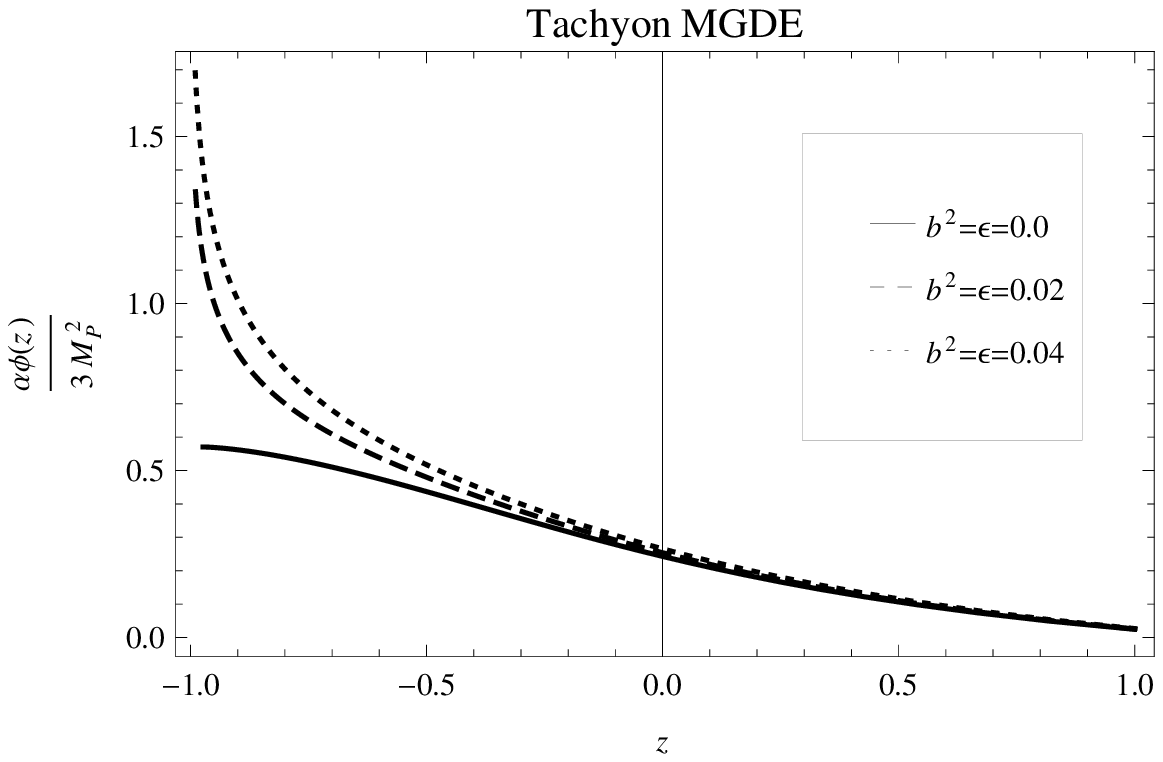}
      \vspace{4.7cm}
      \caption[]{Same as Fig. \ref{PhiT-z-b2} for different coupling $b^2$ and viscosity $\epsilon$ constants.}
         \label{PhiT-z-b2e}
   \end{figure}

\begin{figure}
\includegraphics{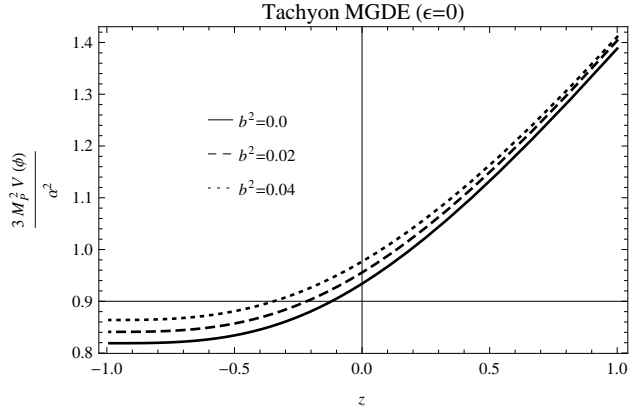}
      \vspace{4.7cm}
      \caption[]{Tachyon MGDE potential, Eq. (\ref{VphiT}), versus redshift for different coupling constants $b^2$ with $\epsilon=0$.
      Auxiliary parameter as in Fig. \ref{PhiQ-z-b2}.}
         \label{VT-z-b2}
   \end{figure}

\begin{figure}
\includegraphics{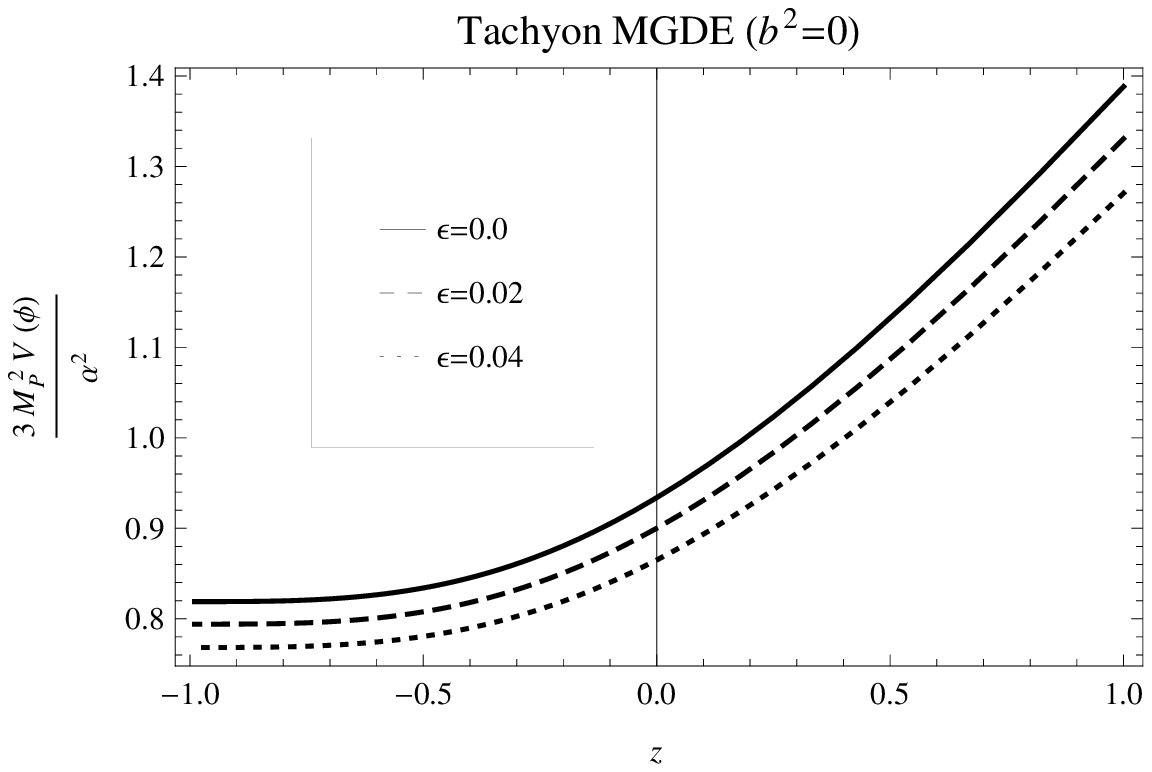}
      \vspace{4.7cm}
      \caption[]{Same as Fig. \ref{VT-z-b2} for different viscosity constants $\epsilon$ with $b^2=0$.}
         \label{VT-z-e}
   \end{figure}

\begin{figure}
\includegraphics{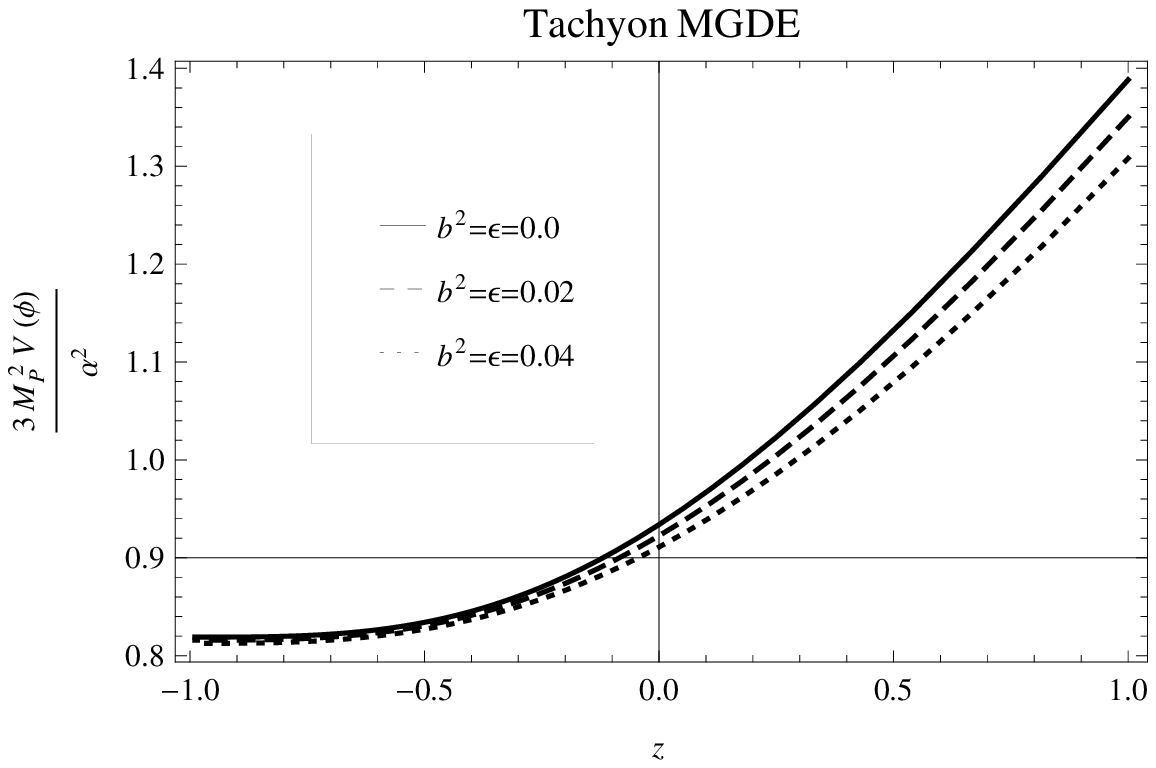}
      \vspace{4.7cm}
      \caption[]{Same as Fig. \ref{VT-z-b2} for different coupling $b^2$ and viscosity $\epsilon$ constants.}
         \label{VT-z-b2e}
   \end{figure}

The tachyon field has been proposed as the source of DE and may be
described by effective field theory corresponding to some sort of
tachyon condensate with an effective Lagrangian density given by
\cite{Sen,Senb,Senc,Send,Sene,Senf}
\begin{equation}
{\mathcal{L}}=-V(\phi)\sqrt{1+\partial_{\mu}\phi
\partial^{\mu}\phi},
\end{equation}
where $\phi$ is a tachyon scalar field and $V(\phi)$ is a potential
of $\phi$. The energy density and pressure of the tachyon scalar
filed are as follows \cite{Sen,Senb,Senc,Send,Sene,Senf}
\begin{equation}\label{rhoT}
    \rho_{T}=\frac{V(\phi)}{\sqrt{1- \dot{\phi^{2}}}},
\end{equation}
\begin{equation}
p_{T}=-V(\phi)\sqrt{1- \dot{\phi^{2}}}.
\end{equation}
The EoS parameter of the tachyon filed reads
\begin{equation}\label{omegaT}
\omega_{T}=\frac{p_{T}}{\rho_{T}}=\dot{\phi^{2}}-1.
\end{equation}
From Eqs. (\ref{rho}) and (\ref{rhoT}), $\rho_{D}=\rho_{T}$ gives
the kinetic energy term
\begin{equation}\label{phidot2}
\dot{\phi^{2}}=\frac{\Omega_{D}-\gamma-\frac{\Omega_{k}}{3}+\frac{\gamma-1}{\Omega_{D}}\left(1+\frac{\Omega_{k}}{3}\right)+2b^{2}
    \left(\frac{1+\Omega_{k}}{\Omega_{D}}\right)}{\Omega_{D}-\gamma-1}
    +3\epsilon.
\end{equation}
Also using Eqs. (\ref{omega}) and (\ref{omegaT}),
$\omega_{D}=\omega_{T}$ yields the tachyon potential
\begin{equation}\label{VphiT}
   V(\phi)=\frac{\alpha^{2}\Omega_{D}}{3M^{2}_{P}(\Omega_{D}+\gamma-1)^{2}}
   \left[\frac{1-\frac{\Omega_{k}}{3}+\frac{\gamma-1}{\Omega_{D}}\left(1+\frac{\Omega_{k}}{3}\right)+2b^{2}
    \left(\frac{1+\Omega_{k}}{\Omega_{D}}\right)}{1+\gamma-\Omega_{D}}-3\epsilon\right]^{1/2}.
\end{equation}
Integrating Eq. (\ref{phidot2}) with respect to $a$ and using
(\ref{OmegaDMGDE}) gives the evolutionary form of the modified ghost
tachyon scalar field as
\begin{eqnarray}\label{e}
  \phi(a)-\phi(a_i)=\frac{3M_{P}^{2}}{\alpha}\int^{a}_{a_i}(\Omega_{D}+\gamma-1)
  ~~~~~~~~~~~~~~~~~~~~~~~~~~~~~~~~~~~~~~~~~~~~~~~~~~~\nonumber\\\times\left[\frac{\Omega_{D}-\gamma-\frac{\Omega_{k}}{3}+\frac{\gamma-1}{\Omega_{D}}\left(1+\frac{\Omega_{k}}{3}\right)+2b^{2}
    \left(\frac{1+\Omega_{k}}{\Omega_{D}}\right)}{\Omega_{D}-\gamma-1}
    +3\epsilon\right]^{\frac{1}{2}}
~\frac{{\rm d}a}{a}.\label{phiT}
\end{eqnarray}
In Figs. \ref{PhiT-z-b2}, \ref{PhiT-z-e} and \ref{PhiT-z-b2e} we
plot the variation of $\phi$, Eq. (\ref{phiT}), versus redshift for
different $b^2$ and $\epsilon$. Also the evolution of tachyon MGDE
potential, Eq. (\ref{VphiT}), is shown in Figs. \ref{VT-z-b2},
\ref{VT-z-e} and \ref{VT-z-b2e}. Figures \ref{PhiT-z-b2} to
\ref{VT-z-b2e} clarify that the scalar field and potential of the
tachyon MGDE model behave like the quintessence one (see Figs.
\ref{PhiQ-z-b2} to \ref{VQ-z-b2e}).

\subsection{K-essence MGDE}

\begin{figure}
\includegraphics{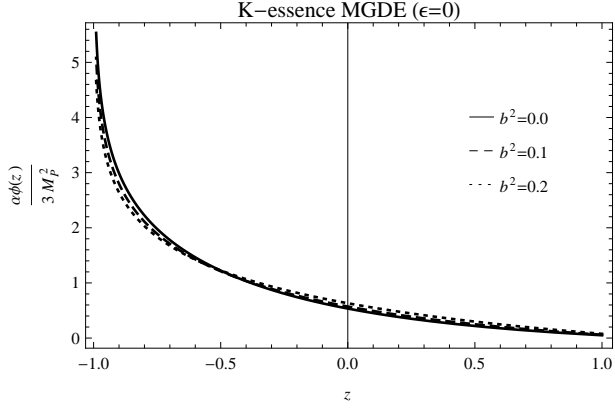}
      \vspace{4.7cm}
      \caption[]{K-essence MGDE scalar field, Eq. (\ref{phiK}), versus redshift for different coupling constants $b^2$
      with $\epsilon=0$. Auxiliary parameters as in Fig. \ref{PhiQ-z-b2}.}
         \label{PhiK-z-b2}
   \end{figure}

\begin{figure}
\includegraphics{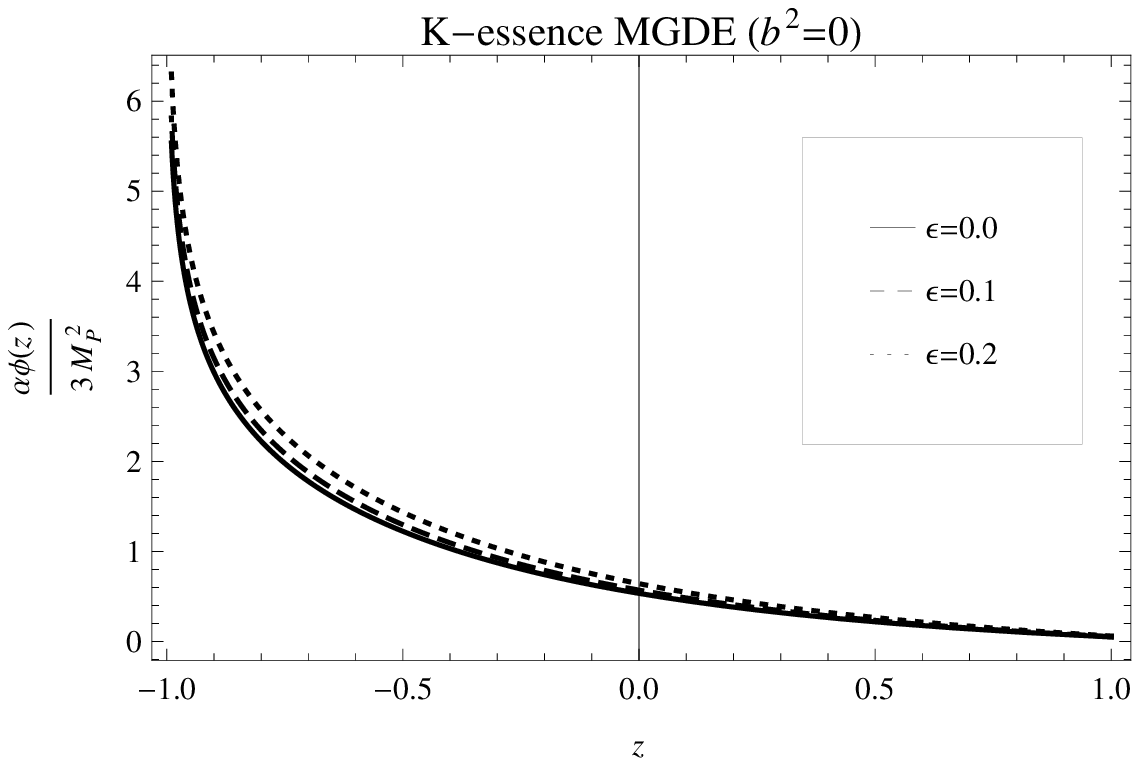}
      \vspace{4.7cm}
      \caption[]{Same as Fig. \ref{PhiK-z-b2} for different viscosity constants $\epsilon$ with $b^2=0$.}
         \label{PhiK-z-e}
   \end{figure}

\begin{figure}
\includegraphics{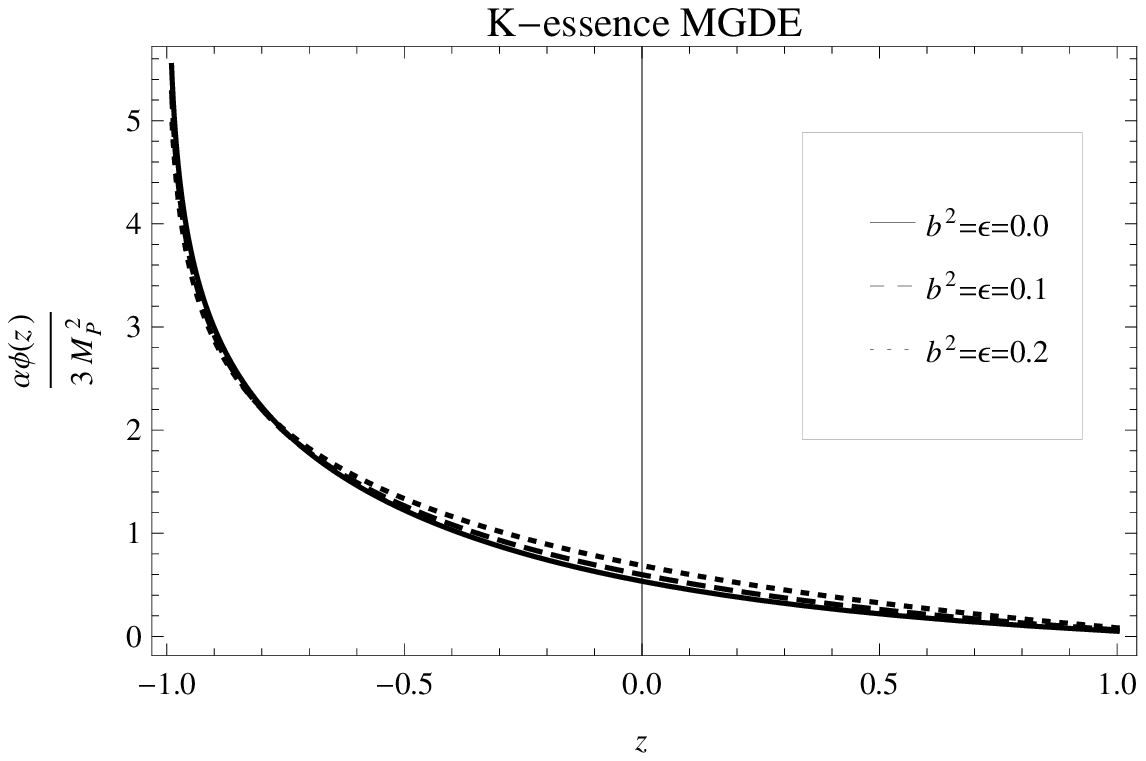}
      \vspace{4.7cm}
      \caption[]{Same as Fig. \ref{PhiK-z-b2} for different coupling $b^2$ and viscosity $\epsilon$ constants.}
         \label{PhiK-z-b2e}
   \end{figure}

The scalar field model known as K-essence is also used to explain
the DE. It is well known that K-essence scenarios have
attractor-like dynamics, and therefore avoid the fine tuning of the
initial conditions for the scalar field
\cite{Chiba,Chibab,Chibac,Picon3,Picon3b}. This kind of models is
characterized by non-canonical kinetic energy terms, and are
described by a general scalar field action which is a function of
$\phi$ and $\chi=\dot{\phi}^2/2$ and is given by
\cite{Chiba,Chibab,Chibac,Picon3,Picon3b}
\begin{equation}
S=\int {\rm d} ^{4}x\sqrt{-{\rm g}}~p(\phi,\chi),\label{action}
\end{equation}
where the Lagrangian density $p(\phi,\chi)$ corresponds to a
pressure with non-canonical kinetic terms as
\begin{equation}
p_K(\phi,\chi)=f(\phi)(-\chi+\chi^{2}),\label{pk}
\end{equation}
and the energy density of the K-essence field $\phi$ is
\begin{equation}
\rho_K(\phi,\chi)=f(\phi)(-\chi+3\chi^{2}).\label{rhok}
\end{equation}
One of the motivations to consider this type of Lagrangian
originates from considering low energy effective string theory in
the presence of a high order derivative terms. The EoS parameter of
the K-essence scalar field is obtained as
\begin{equation}\label{omegaK}
    \omega_{K}=\frac{p_K}{\rho_K}=\frac{\chi-1}{3\chi-1}.
\end{equation}
Following Eqs. (\ref{omega}) and (\ref{omegaK}),
$\omega_{D}=\omega_{K}$ gives
\begin{equation}\label{chiK}
\chi=\frac{2+\gamma-\frac{\Omega_k}{3}+\frac{\gamma-1}{\Omega_{D}}\Big(1+\frac{\Omega_k}{3}\Big)
+2b^2\Big(\frac{1+\Omega_k}{\Omega_{D}}\Big)-\Omega_{D}+3\epsilon(\Omega_{D}-\gamma-1)}
{4+\gamma-\Omega_k+\frac{\gamma-1}{\Omega_{D}}(3+\Omega_k)+6b^2\Big(\frac{1+\Omega_k}{\Omega_{D}}\Big)
-\Omega_{D}+9\epsilon(\Omega_{D}-\gamma-1)}.
\end{equation}
Integrating this with respect to $a$ gives the scalar field of the
K-essence MGDE as
\begin{eqnarray}
  \phi(a)-\phi(a_i)= \frac{3M_{P}^{2}}{\alpha}\int^{a}_{a_i}(\Omega_{D}+\gamma-1)
~~~~~~~~~~~~~~~~~~~~~~~~~~~~~~~~~~~~~~~~~~~~~~~~~~~~~~~~~~~~~~~~~~\nonumber\\\times\
\left[\frac{4+2\gamma-\frac{2\Omega_k}{3}+2\Big(\frac{\gamma-1}{\Omega_{D}}\Big)\Big(1+\frac{\Omega_k}{3}\Big)
+4b^2\Big(\frac{1+\Omega_k}{\Omega_{D}}\Big)-2\Omega_{D}+6\epsilon(\Omega_{D}-\gamma-1)}
{4+\gamma-\Omega_k+\Big(\frac{\gamma-1}{\Omega_{D}}\Big)(3+\Omega_k)+6b^2\Big(\frac{1+\Omega_k}{\Omega_{D}}\Big)
-\Omega_{D}+9\epsilon(\Omega_{D}-\gamma-1)}\right]^{\frac{1}{2}}
~\frac{{\rm d}a}{a}.\label{phiK}
\end{eqnarray}
The evolutionary behavior of this field for different $b^2$ and
$\epsilon$ is plotted in Figs. \ref{PhiK-z-b2}, \ref{PhiK-z-e} and
\ref{PhiK-z-b2e}. Figures present that the K-essence MGDE scalar
field increases during history of the universe.

\subsection{Dilaton MGDE}

The dilaton scalar field model is also an interesting attempt to
explain the origin of DE using string theory. This model appears
from a four dimensional effective low energy string action and
includes higher order kinetic corrections to the tree level action
in low energy effective string theory \cite{Gasperini,Gasperinib}.

\begin{figure}
\includegraphics{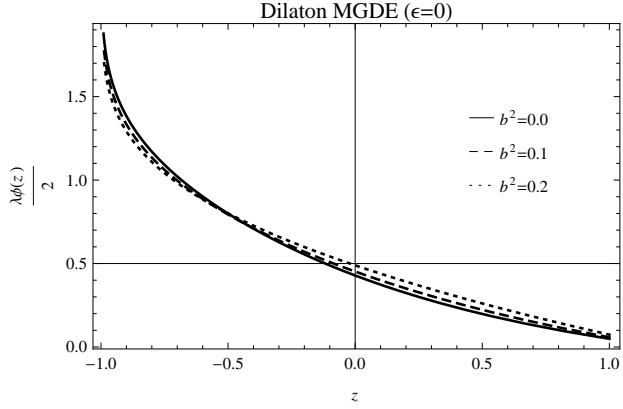}
      \vspace{4.7cm}
      \caption[]{Dilaton MGDE scalar field, Eq. (\ref{phiD}), versus redshift for different coupling constants
      $b^2$ with $\epsilon=0$.  Auxiliary parameters are $\Omega_{k_0}=0.01$, $\Omega_{D_0}=0.76$, $\gamma=1.105$ \cite{Cai2}, $\phi(1)=0$
      and $\delta=\frac{3M_p^2\lambda}{2\alpha\sqrt{c}}=1$.}
         \label{PhiD-z-b2}
   \end{figure}

\begin{figure}
\includegraphics{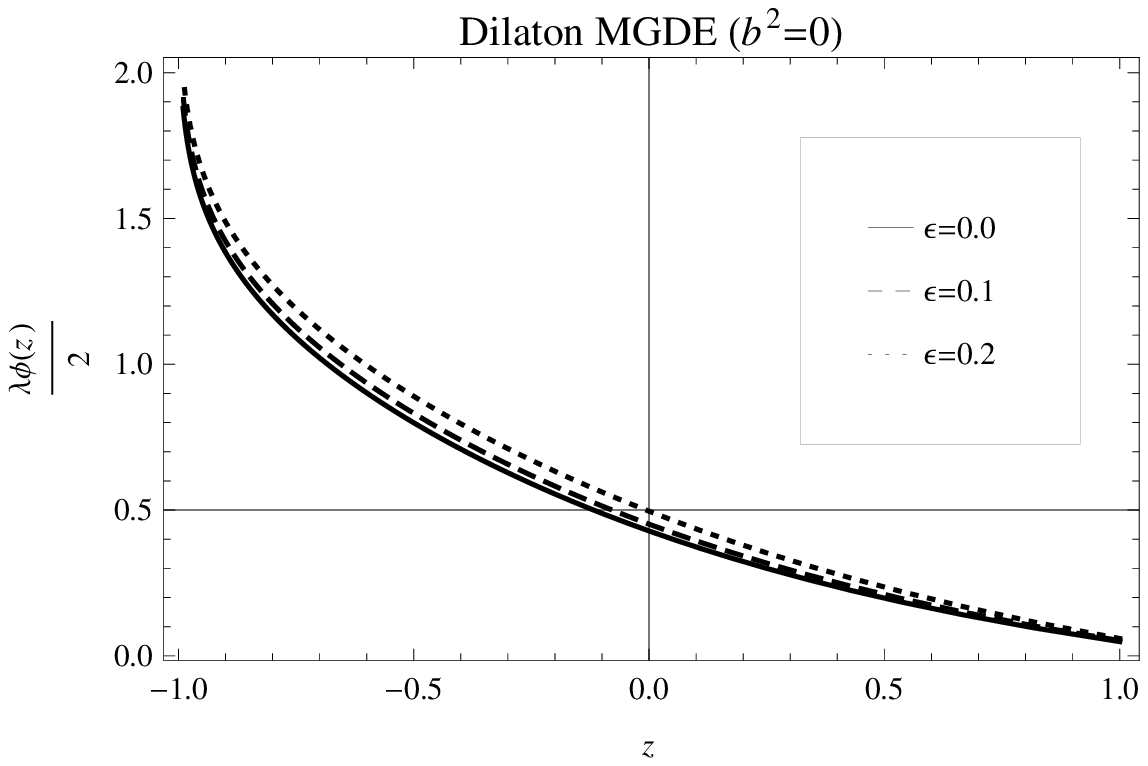}
      \vspace{4.7cm}
      \caption[]{Same as Fig. \ref{PhiD-z-b2} for different viscosity constants $\epsilon$ with $b^2=0$.}
         \label{PhiD-z-e}
   \end{figure}

\begin{figure}
\includegraphics{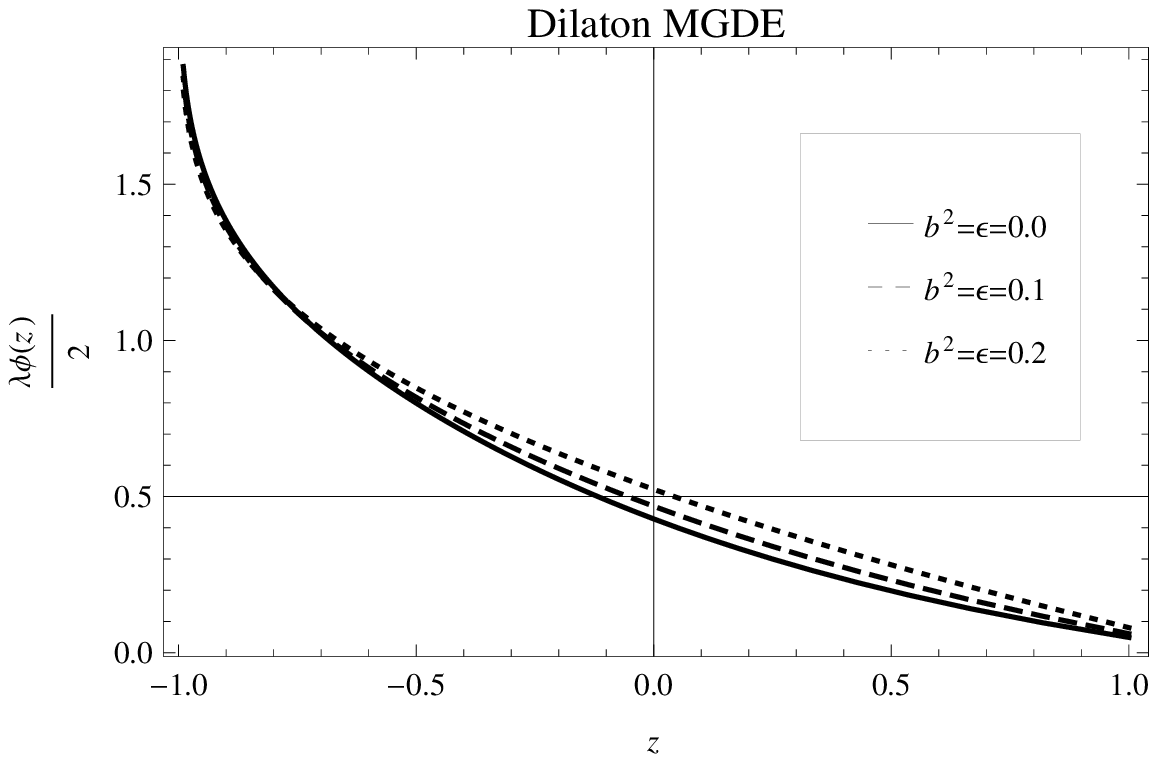}
      \vspace{4.7cm}
           \caption[]{Same as Fig. \ref{PhiD-z-b2} for different coupling $b^2$ and viscosity $\epsilon$ constants.}
         \label{PhiD-z-b2e}
   \end{figure}

For the dilaton scalar field, the pressure and energy density take
the forms \cite{Gasperini,Gasperinib}
\begin{equation}
    p_{D}=-\chi+ce^{\lambda\phi}\chi^{2},\label{pD}
\end{equation}
\begin{equation}\label{rhoD}
    \rho_{D}=-\chi+3ce^{\lambda\phi}\chi^{2}.
\end{equation}
Here $c$ and $\lambda$ are two positive constants and
$\chi=\dot{\phi}^2/2$ is the dilaton kinetic energy. Therefore, the
dilaton EoS parameter reads
\begin{equation}\label{omegaD}
    \omega_{D}=\frac{p_{D}}{
    \rho_{D}}=\frac{-1+ce^{\lambda\phi}\chi}{-1+3ce^{\lambda\phi}\chi}.
\end{equation}
Identifying Eq. (\ref{omega}) with (\ref{omegaD}) reduces to
\begin{equation}
    ce^{\lambda\phi}\chi=\frac{2+\gamma-\frac{\Omega_k}{3}+\frac{\gamma-1}{\Omega_{D}}\Big(1+\frac{\Omega_k}{3}\Big)
+2b^2\Big(\frac{1+\Omega_k}{\Omega_{D}}\Big)-\Omega_{D}+3\epsilon(\Omega_{D}-\gamma-1)}
{4+\gamma-\Omega_k+\frac{\gamma-1}{\Omega_{D}}(3+\Omega_k)+6b^2\Big(\frac{1+\Omega_k}{\Omega_{D}}\Big)
-\Omega_{D}+9\epsilon(\Omega_{D}-\gamma-1)}.
\end{equation}
Using $\chi=\dot{\phi}^2/2$ one can take the integral of above
equation with respect to $a$. The result yields
\begin{eqnarray}
  \phi(a)=\frac{2}{\lambda}\ln\left\{e^\frac{\lambda\phi(a_i)}{2}+\delta\int^{a}_{a_i}(\Omega_{D}+\gamma-1)
\right.~~~~~~~~~~~~~~~~~~~~~~~~~~~~~~~~~~~~~~~~~~~~~~~~~~~~~~~~~~~~~~~~~~\nonumber\\\left.\times\
\left[\frac{4+2\gamma-\frac{2\Omega_k}{3}+2\Big(\frac{\gamma-1}{\Omega_{D}}\Big)\Big(1+\frac{\Omega_k}{3}\Big)
+4b^2\Big(\frac{1+\Omega_k}{\Omega_{D}}\Big)-2\Omega_{D}+6\epsilon(\Omega_{D}-\gamma-1)}
{4+\gamma-\Omega_k+\Big(\frac{\gamma-1}{\Omega_{D}}\Big)(3+\Omega_k)+6b^2\Big(\frac{1+\Omega_k}{\Omega_{D}}\Big)
-\Omega_{D}+9\epsilon(\Omega_{D}-\gamma-1)}\right]^{\frac{1}{2}}
~\frac{{\rm d}a}{a}\right\},\label{phiD}
\end{eqnarray}
where $\delta=\frac{3M_p^2\lambda}{2\alpha\sqrt{c}}$. In Figs.
\ref{PhiD-z-b2}, \ref{PhiD-z-e} and \ref{PhiD-z-b2e}, variation of
the dilaton MGDE scalar field, Eq. (\ref{phiD}), is illustrated for
different $b^2$ and $\epsilon$. Figures clear that the scalar field
of the dilation MGDE model behaves like the K-essence one (see Figs.
\ref{PhiK-z-b2}, \ref{PhiK-z-e} and \ref{PhiK-z-b2e}).
\section{Conclusions}
Here we studied the Veneziano MGDE model in the framework of
Einstein's gravity. We considered a spatially non-flat FRW universe
filled with interacting viscous MGDE and DM. We derived a
differential equation governing the evolution of the MGDE density
parameter and solved it numerically. We also obtained the EoS
parameter of the interacting viscous MGDE and the deceleration
parameter of the universe. Moreover, using the linear perturbation
theory we investigated the evolution of growth of structure in our
model. Furthermore, using a correspondence between the interacting
viscous MGDE and quintessence, tachyon, K-essence and dilaton scalar
field models of DE we reconstructed the dynamics $\phi$ and
potentials $V(\phi)$ of the aforementioned scalar field models
according to the evolution of the MGDE density. Our numerical
results show the following.

(i) The MGDE density parameter $\Omega_D$ for a given coupling
constant $b^2$, increases when the time increases. The evolutionary
behavior of $\Omega_D$ is independent of viscosity.

(ii) The variation of the deceleration parameter $q$ shows that the
universe transitions from an early matter dominant epoch to the de
Sitter era in the future, as expected. Also $q$ like $\Omega_D$ does
not depend on viscosity.

(iii) The EoS parameter $\omega_D$ of the MGDE model in the absence
of viscosity ($\epsilon=0$), varies from the quintessence phase to
the phantom regime. Whereas in the absence of interaction ($b^2=0$),
it behaves like the freezing quintessence model. The interaction and
viscosity have opposite effects on $\omega_D$.

(iv) The evolution of DM density perturbation $\delta_{\rm m}$ in
the presence of interaction shows a suppression of growth of
structure. Since the DM density is lower in the past relative to the
non-interacting case, leading to a suppression of growth of
structure. Here the viscosity constant $\epsilon$ does not affect
$\delta_{\rm m}$, because the DM has not the viscosity properties in
our model.

(v) The quintessence and tachyon MGDE scalar fields for a given
$b^2$ or $\epsilon$, increase during history of the universe. For a
given redshift, they decrease and increase with increasing $b^2$ and
$\epsilon$, respectively. The potentials of the quintessence and
tachyon MGDE models for a given $b^2$ or $\epsilon$, decrease with
increasing the time. The interaction and viscosity have opposite
effects on both the scalar fields and potentials of the
aforementioned models.

(vi) The K-essence and dilaton MGDE scalar fields for a given $b^2$
or $\epsilon$ increase during history of the universe.

\subsection*{Acknowledgements}

The authors thank the unknown referees for very valuable comments.
The work of K. Karami has been supported financially by Center for
Excellence in Astronomy \& Astrophysics of Iran (CEAAI-RIAAM), under
research project No. 1/3076.


\clearpage
\clearpage
\clearpage
\clearpage
\clearpage
\clearpage
\clearpage
\clearpage
\clearpage

\end{document}